# User Interface as a Thinging Machine

Sabah Al-Fedaghi

Computer Engineering Department, Kuwait University
P.O. Box 5969 Safat 13060 Kuwait
sabah.alfedaghi@ku.edu.kw

**Abstract.** The availability of interaction devices has raised interest in techniques to support the user interface (UI). A UI specification describes the functions that a system provides to its users by capturing the interface's details and includes possible actions through interaction elements. UI developers of interactive systems have to address multiple sources of heterogeneity, including end users' heterogeneity and variability of the context of use. This paper contributes to the notion of interactivity and interfacing by proposing a methodology for producing engineering-type diagrams of (abstract) machine processes that can specify uniform structure and behavior of systems through a synchronic order of states (stages): creation, release, transfer, receive, and process. As an example, the diagrammatic methodology is applied to conceptualizing space as a machine. The resulting depiction seems suitable for use in designing UIs in certain environments.

## 1 Introduction

The availability of interaction devices has raised interest in techniques to support adaptation of the user interface. In Web applications, the adaptation can take place in the application server, the proxy server, or the client device [1]. Typically, a user interface (UI) specification describes the functions that a system provides to its users by capturing the interface's details and includes possible actions that an end user may complete through interaction elements [1].

UI development, ranging from requirement specifications to software implementation, has required extensive efforts, especially with the spread of new interaction techniques (e.g., vocal and gestural modalities) [2]. Currently, UI developers of interactive systems have to address multiple sources of heterogeneity including [3]:

1. End users' heterogeneity: An interactive system is normally used by various end users. End users differ with respect to their preferences, capabilities, cultures, and levels of experience.

2. Variability of the context of use: In addition to being heterogeneous, the context of use dynamically evolves, calling for plastic UIs (i.e., UIs capable of adapting while preserving human values).

This paper introduces a unifying framework for a diagrammatic description of a user, a system, and the interface that sits between them to provide the context for the notion of *interaction*. The paper examines conceptual models and their applications in UI where the focus is on the "patterning aspect of cognition" (i.e., object-oriented) or recurring templates used in thinking. It is proposed that representations of thinking activity be based on the flows of things (to be defined later) using abstract machines formed by stages (states) occurring sequentially in a flow. According to such an approach, a user's "thought machine" forms a train of thought that excludes other modes, such as procedural and object-oriented modes of thinking. The new idea presented here is that modeling is used not only as an external representation of the user's thinking but also as the style of thinking. A thinking style involves how one organizes thoughts and is a "conscious system of interaction." We illustrate the method by remodeling examples from the literature. It seems to have merits that deserve further development.

Since the suggested modeling technique is not yet widely known, and for the sake of a self-contained paper, the next section briefly reviews the proposed diagrammatic language, called the Thinging Machine (TM) model, which forms the foundation of this paper's theoretical development. This diagrammatic language has been adapted for several applications [4-11]; however, the example given here is a new contribution.

## 2 Thinging Machines (TM)

This model is a diagrammatic schema that uses *things* to represent a range of items, (see Figure 1). In general, a machine is thought to be an abstract machine that receives, processes, creates, releases, and transfers things. TM also uses the notions of spheres and subspheres. These are the network environments and relationships of machines and submachines. TM also utilizes the notion of *triggering*. Triggering is denoted in TM diagrams by a dashed arrow.

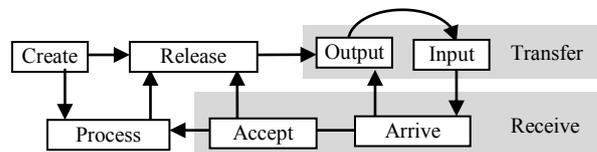

**Figure 1. Thinging machine**

**Example**: According to Dix et al. [12], a dialogue refers to the structure of the interaction or syntactic level of human–computer "conversation." Dialogue graphical notations include state-transition nets (STN), Petri nets, state charts, and flow charts. Figure 2 shows a sample STN, where circles are states and arcs are actions or events.

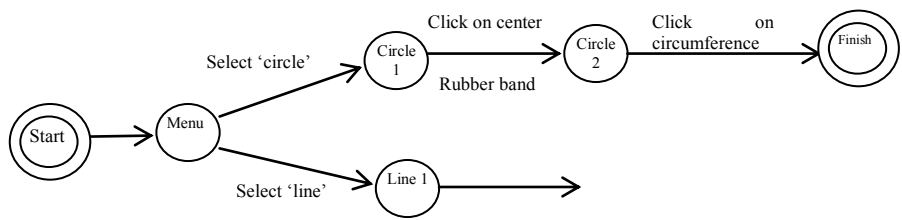

**Figure 2. Sample STN (Re-drawn, partial from Dix et al. [12])**

### 2.1 TM static description

Figure 3 shows the corresponding TM representation drawn according to our understanding of the given example.

First, the user requests to login (circle 1). The request flows to the system (2), where it is processed (3). If the user request is approved, the flow is triggered (4), releasing the Menu (5), which flows to the user to be processed (6). Processing the menu results in selecting "circle" or "line," (7), which flows to be processed (8). If the selection is "circle," then it triggers the release of the circle shape (9) to the user (10) to be processed. There, the user has three actions to choose from: *clicking on the center* (11), *rubber band* (12), or *clicking on the circumference* (13). If the user *clicks on the center* (14), then this action flows to the system to be processed (15), resulting in the processing of the circle, (16) which changes accordingly, and releasing it (9) back to the user (10). A similar process occurs when the user selects *rubber band* (17). It is not clear from the original description what happens when the selection is *click on the circumference* (e.g., printing the resultant circle, storing it, etc.), so we indicate receiving the message without taking an action (19). The rest of process for the selection *line* can be described in a similar way.

We will refer to Figure 3 as the *Diagram* of the system.

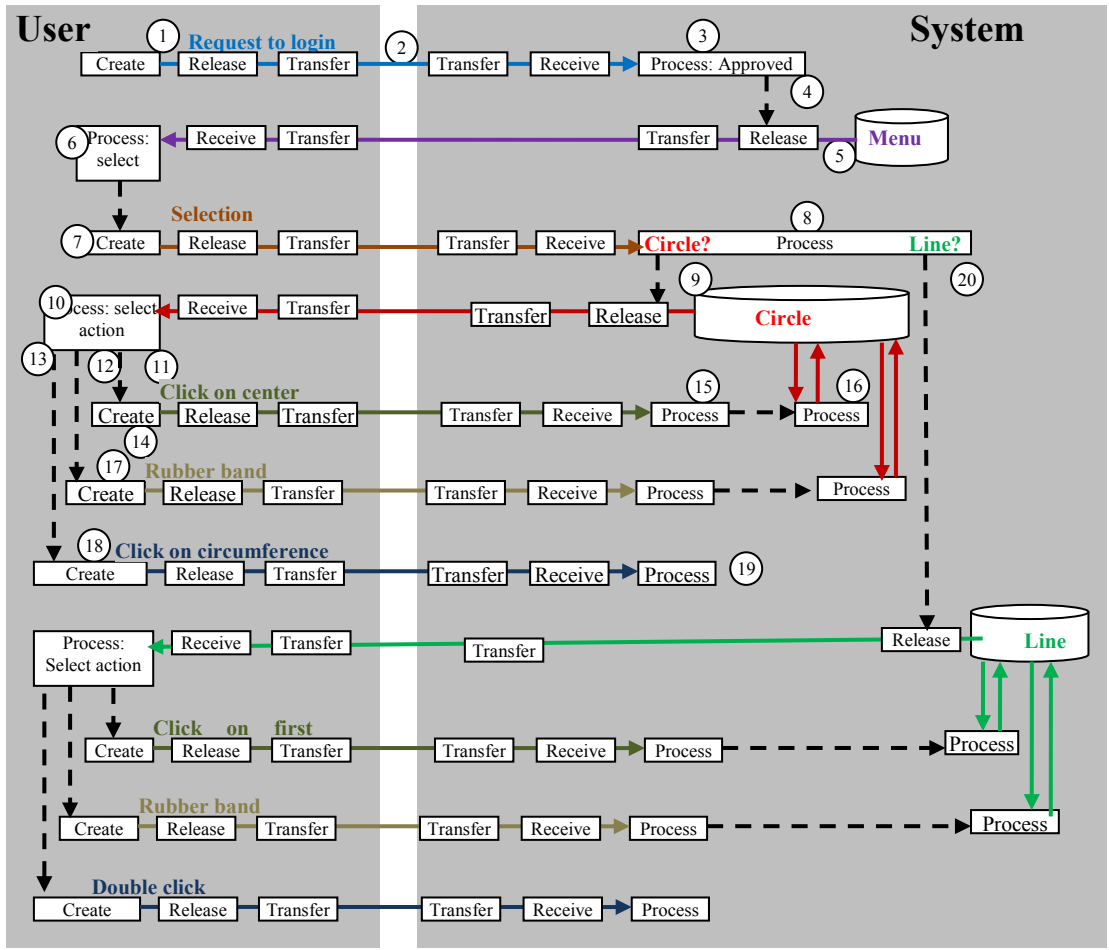

**Figure 3. The TM description of the example.**

## 2.2 TM Behavior description

The specifications of the dynamic behavior can be used in controlling the execution. C*ontrol* here refers to the configuration of facilitating system dynamics (events) (e.g., sequence, parallelism, timing, etc.). It oversees activities "according to specification." *Behavior* involves *things* during *events* *w*hen the *diagram* is acted upon. The chronology of events can be identified by orchestrating these events in their interacting processes. In FM, an event is a *thing* that can be created, processed, released, transferred, and received. An event *sphere* consists of at least the event itself, its time, and its region (sub-diagram of the system's *diagram*). Accordingly, the choreography of execution can be revealed by the arrangement of events. We identify regions of "meaningful" events as sub-diagrams, as follows (see Figure 4; only the regions of the events are shown):

**Event 1 ($E_1$)**: A request to login is created, flows to the system, and is processed.
**Event 2 ($E_2$)**: An approved login sends the menu to the user.
**Event 3 ($E_3$)**: A selection is made that flows to the system.
**Event 4 ($E_4$)**: The selection is *circle*; hence, a circle shape is displayed to the user.
**Event 5 ($E_5$)**: The circle is clicked on the center.
**Event 6 ($E_6$)**: The circle is rubber banded.
**Event 7 ($E_7$)**: The circle is clicked on the circumference
**Event 8 ($E_8$)**: The circle is processed according to "clicked on the center."
**Event 9 ($E_9$)**: The circle is processed according to "rubber band."

Fig. 5 shows the execution sequence of these events.

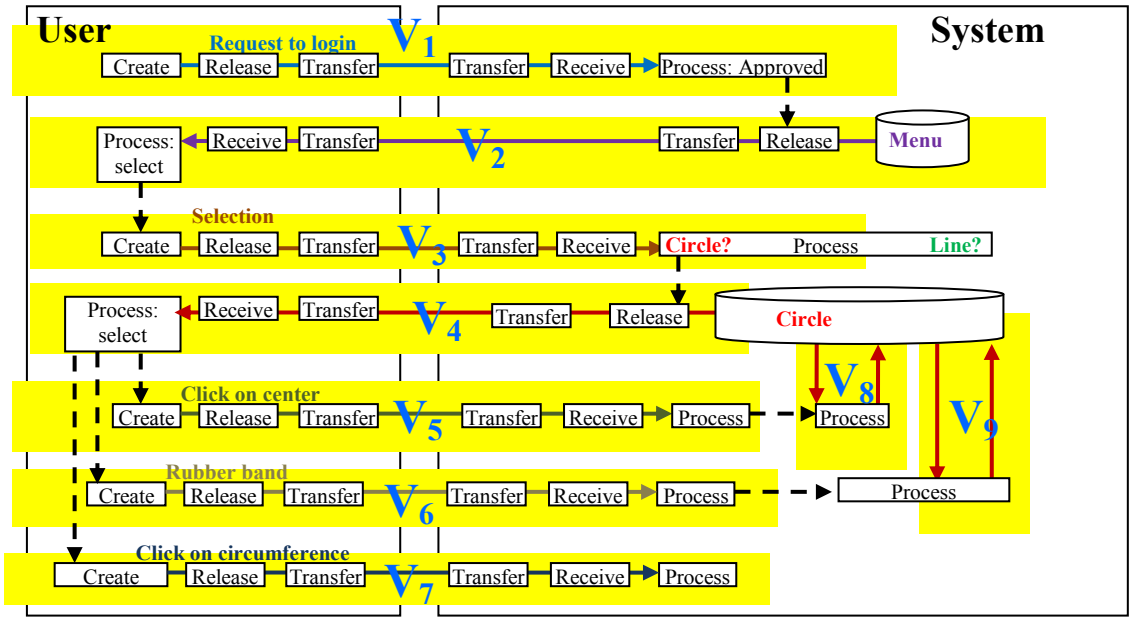

**Figure 4. The TM description of the example**

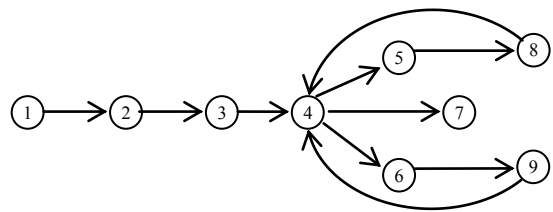

**Figure 5. Events sequence of the example**

## 3  Sample TM Application: Digital Home

   The *home* is becoming more populated by intelligent devices with the ability to communicate information, thus allowing users to access domestic appliances from anywhere using interactive devices [1]. According to W3C [3],

   Digital home refers to a residence with devices that are connected through a computer network. A digital home has a network of consumer electronics, mobile, and PC devices that cooperate transparently. All computing devices and home appliances can be controlled by means of an interactive system. These functionalities are made available through context sensitive user interfaces.

The key functions include authenticating a user, selecting a room, selecting a device inside the room, and inspecting and modifying a selected device's status. Figure 6 shows a partial view of a screenshot of the Digital Home for a targeted context of use, given W3C [3].

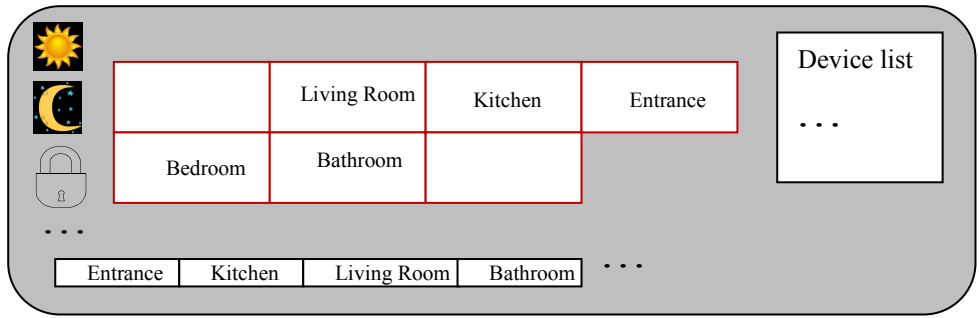

**Figure 6. The desktop version of the Digital Home application (Re-drawn, partial from W3C [3])**

TM can present the interface in this context within the notion of "house as machine" [13] by proposing a *one whole* interface instead of an ad hoc collection of icons. The *one wholeness* is used to describe the structure and dynamic behavior exhibited by a system where, at each step of building, "the wholeness of the structure is preserved [the TM machine] though extended, each time changing without losing the global structure" [14]. The TM diagram is a machine with an "underlying process whose steps are wholeness-extending-transformations, where each [such] transformation operates on one wholeness [machine] to produce another wholeness [machine] [15].

In this example, it can be noted that the problem is a question of spaces, their contents, and flows, which can be conceptualized as things and machines. According to Le Corbusier [16], a house is a machine for living in. A machine, in this sense, indicates a system of activities and actions: "We use a machine, or the drawing of a machine, to symbolize a particular action of the machine. For instance, we give someone such a drawing and assume that he will derive the movement of the parts from it" [17].

For this example, the TM diagrammatic methodology is based on the conceptual model of space as a synchronic order of states (stages): creation, release, transfer, receive, and process. It views space in terms of dynamic content with stages and connections (flows) to create a representation of the flows in a home. Figure 7 shows an TM representation of the problem under consideration. A hall is added to reflect reality. In FM, a space is a machine. It has dynamic content. Activities (stages) and their connections (flows) in space establish a flow system (machine) representation. As an interface, a user has to think this way to achieve uniformity of thinking.

We propose flow-based thinking by the interface users instead of other types of thinking, such as categories and types that trigger other types (e.g., object-oriented). In forcing this FM-based thinking, we isolate and cut nature up into *things that flow in their spheres* (environment). Accordingly, the proposed FM-based user interface is suitable for those users who think this way, as shown in Figure 8.

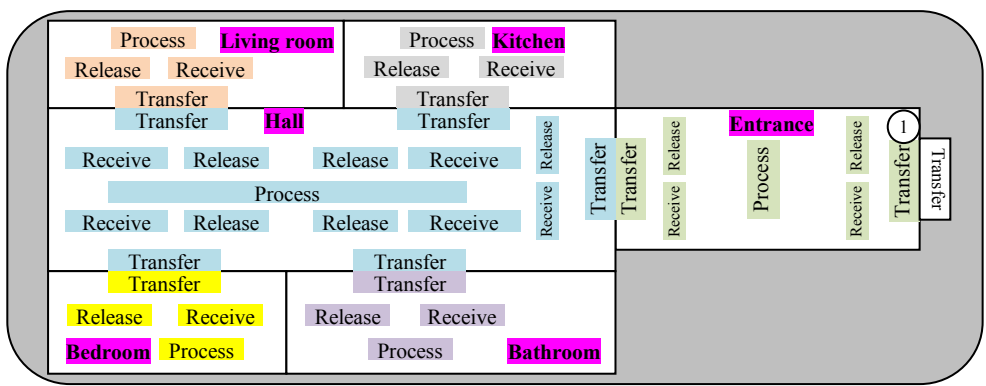

**Figure 7. General Interface shown to the user**

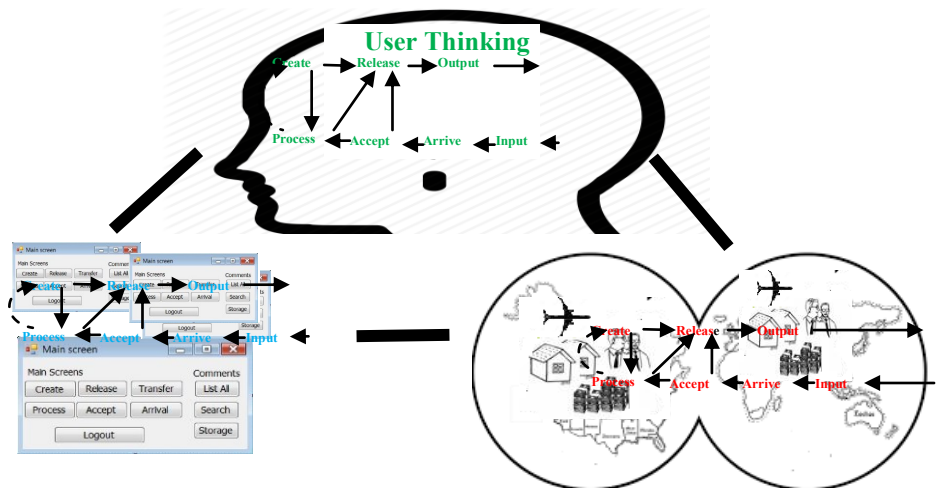

**Figure 8. FM-based world perception, thinking and user interface**

Figure 7 reflects a general view of the house as a machine to be displayed to the user. Accordingly, the user virtually "enters" the house by clicking on the *Transfer* (circle 1). As a result, the screen shown in Figure 9 is displayed (2). It includes the door (3) and light (4). The user can check if the door is *Open* or *Closed* by clicking on its *State* (4). If it is *Closed*, the user cannot progress any further unless he or she opens it using a password. The sub-widows for opening the door are not shown. Assuming that the door is open, the user can turn on the door's light. Progressing further, the user "enters" the house by clicking on Receive (5) in Figure 9 to go to the *entrance area beyond* (6), as displayed in Figure 10. The user moves virtually from one part of the house to another, flowing through TM stages where each stage shows what is in the area as a curtain opens and displays what is behind it.

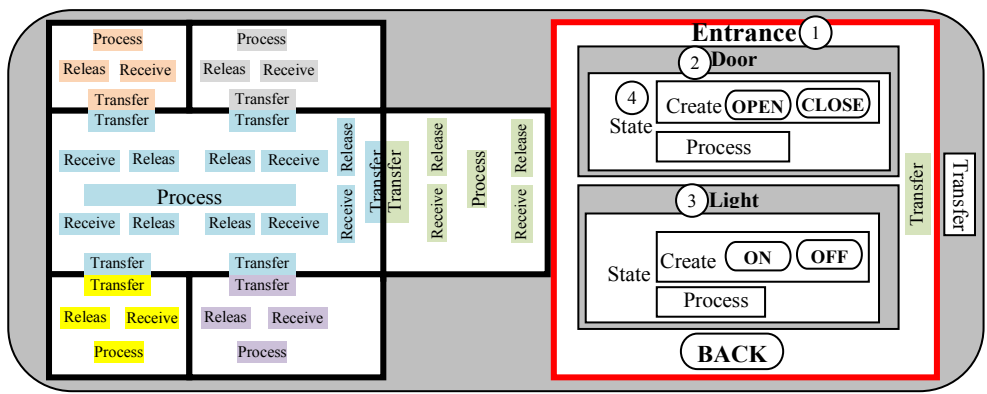

**Figure 9. The entrance door area is displayed**

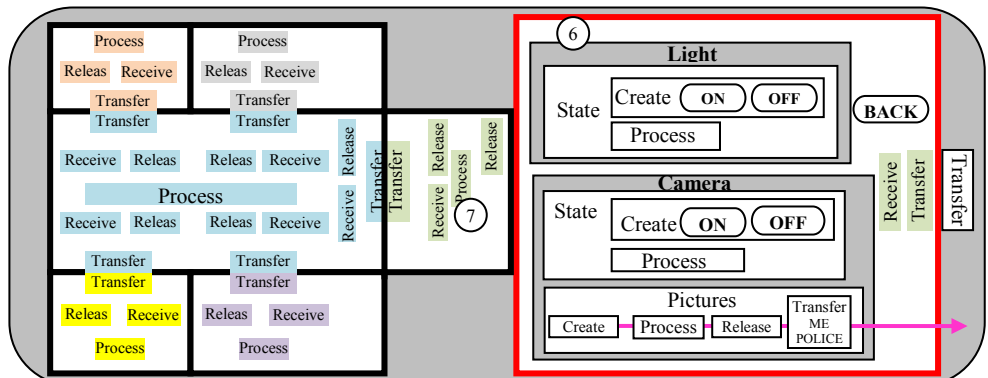

**Figure 10. The entrance is displayed**

Now the user is virtually in the *Entrance area beyond the door*, where there is a *Light* and a *Camera* that can be controlled in an obvious way. Then the user leaves the *Receive* stage by clicking on *Process* (7) in the *Entrance beyond the door*. This action causes the display of possible processes (8 – Figure 11) in the area, such as cleaning, disinfection, coloring it through "ordering," or controlling some person there, or a robot.

Now we jump the view and assume that there is an elderly person in the bedroom that needs to be monitored. Figure 12 displays the view when the user reaches the bedroom (9). We assume that the sensor in the bed signals that if there is an elderly person there (10), assuming that the elderly person is not found in the bed. It is possible then to ask the system to display the elder person's last activities, as shown in Figure 13. The figure (through sensors) tells us that he/she left the bedroom through the hall to the bathroom, where he opened the door (*Hall.Release*, *Hall.Transfer*) and moved through the door, but he/she had never been *Received* in the bathroom. One possibility of this situation is that he/she passed out, as when having a heart attack, and fell in the door area.

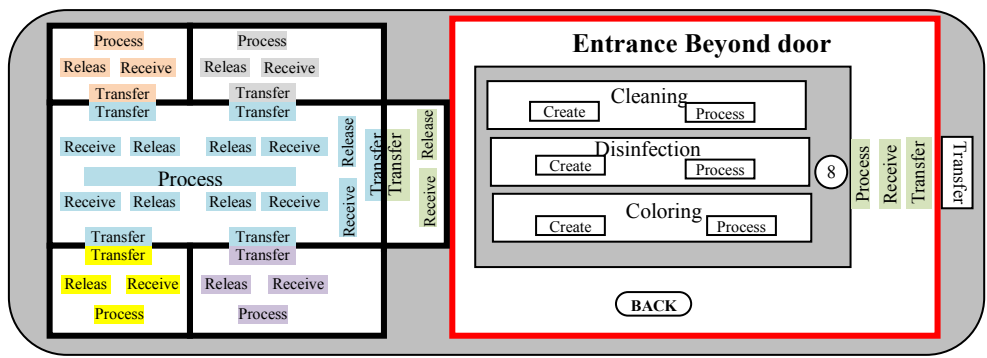

**Figure 11. Possible processing of the entrance area**

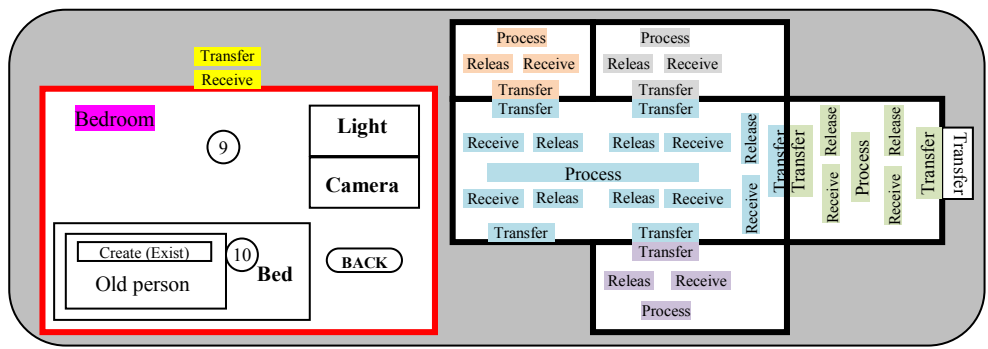

**Figure 12. The screen while viewing the bedroom**

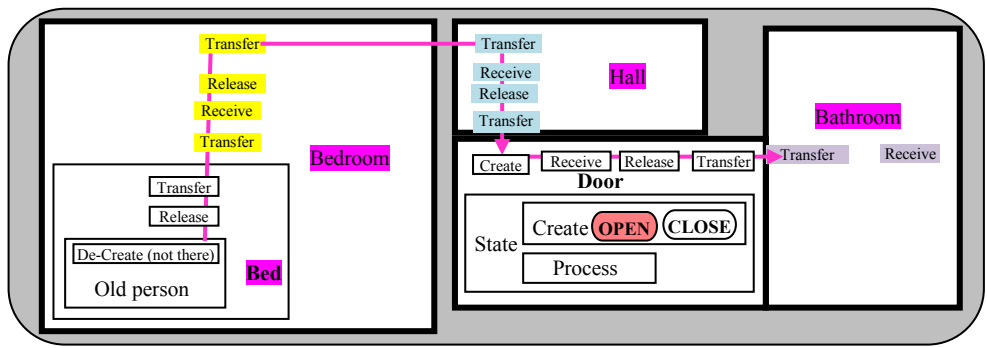

**Figure 13. The screen of the activities of the elderly person**

## 4 Conclusion

This paper has explored applying a new diagrammatic language in the context of UI in place of or as a complement to the ad hoc graphs utilized currently in this area. Many issues remain to be clarified; however, preliminary results demonstrate the approach's potential feasibility.